\documentclass[preprint,showpacs,preprintnumbers,amsmath,amssymb]{revtex4}

\usepackage{graphicx}
\usepackage{dcolumn}
\usepackage{bm}

\begin{document}
\preprint{}
\title{Relativistic predictions of spin observables 
for exclusive proton knockout reactions}
\author{G. C. Hillhouse$^{1,\, 2,\,}$\footnote{\uppercase{E}lectronic mail: gch@sun.ac.za}, 
J. Mano$^{3}$, S. M. Wyngaardt$^{4}$,  B. I. S. van der Ventel$^{1}$,
T. Noro$^{5}$ and K. Hatanaka$^{2}$, 
}  
\affiliation{$^{1}$Department of Physics, University 
of Stellenbosch, Private Bag X1, Matieland 7602, South Africa\\
$^{2}$Research Center for Nuclear Physics, Osaka
University, Ibaraki, Osaka 567-0047, Japan\\
$^{3}$Department of Electrical Engineering and 
Computer Science, Osaka Prefectural College of 
Technology, Osaka 572-8572, Japan\\
$^{4}$Department of Physics, University of the
Western Cape, Private Bag X17, Bellville 7535, South Africa\\
$^{5}$Department of Physics, Kyushu University, 
Fukuoka, 812-8581, Japan\\
}
\date{\today}

\begin{abstract}
Within the framework of the relativistic distorted wave impulse approximation (DWIA), 
we investigate the sensitivity of complete sets of polarization transfer observables 
($\mbox{P},\, \mbox{A}_{y},\, \mbox{D}_{n n},\, \mbox{D}_{s' s},\, \mbox{D}_{\ell' \ell},\, 
\mbox{D}_{s' \ell},\ \mbox{and}\ \mbox{D}_{\ell' s}$), for exclusive proton knockout 
from the 3s$_{1/2}$, 2d$_{3/2}$ and  2d$_{5/2}$ states in $^{208}$Pb, at an incident 
laboratory kinetic energy of 202 MeV, and for coincident coplanar scattering angles 
($28.0^{\circ}$, $-54.6^{\circ}$), to different distorting optical potentials, 
finite-range (FR) versus zero-range (ZR) approximations to the DWIA, as well as 
medium-modified meson-nucleon coupling constants and meson masses. Results are also 
compared to the nonrelativistic DWIA predictions based on the Schr\"{o}dinger equation.

For knockout from the 3s$_{1/2}$ state, A$_{y}$, D$_{n n}$ and D$_{s' s}$ exhibit large 
differences between Dirac and Schr\"{o}dinger-based DWIA models. On the other hand, for 
knockout from the 2d$_{3/2}$ state, the most sensitive observables to the latter are 
D$_{n n}$, D$_{\ell ' s}$ and D$_{\ell ' \ell}$, whereas the corresponding observables for 
2d$_{5/2}$ knockout are D$_{s' s}$ and D$_{\ell ' s}$. The most sensitive observables
to ZR versus FR approximations to the DWIA are the induced polarization (P), D$_{\ell' s}$  
and D$_{s' \ell}$ for knockout from the 3s$_{1/2}$, 2d$_{3/2}$ and 2d$_{5/2}$ states,
respectively.

Although polarization transfer observables are relatively insensitive to different global Dirac optical 
potential parameter sets, distorting optical potentials are crucial for describing the 
oscillatory behavior of spin observables. In addition, it is seen that A$_{y}$, P and D$_{s' s}$ 
are very sensitive to reductions in the meson-coupling constants and meson masses by the nuclear 
medium for proton knockout from all three states. 
\end{abstract}

\maketitle

\section{\label{sec:introduction}Introduction}
It is now well established that spin observables are more appropriate
than unpolarized cross sections for discriminating between different 
physical processes partaking in nuclear reactions \cite{Sp}. Different spin 
observables usually exhibit selective sensitivity to different 
physical effects and, hence, in order to test the validity of a
theoretical model, it is advisable to measure as many independent 
spin observables as possible or, at the very least one needs to identify
(via model predictions) specific observables which can potentially address 
the physical problem of interest.

One of the most challenging problems in nuclear physics is to understand how 
the properties of the strong interaction are modified inside nuclear matter.
Various theoretical models \cite{Se86,Br91,Fu92} predict the modification of 
meson-nucleon coupling constants as well as nucleon and meson masses by normal 
nuclear matter. At present there is no overwhelming experimental evidence supporting 
these predictions. However, we believe that exclusive ($\vec{p},2\vec{p}\,$) reactions - 
whereby an incident polarized proton knocks out a bound proton from a specific 
orbital in the nucleus and the two scattered protons, one of which is polarized, are 
detected in coincidence - are ideally suited for studying the behavior of the 
nucleon-nucleon (NN) interaction in the nuclear medium. By exploiting the 
discriminatory nature of independent spin observables for the knockout protons 
from deep to low-lying single particle states in nuclei, one can in principle
extract information about the density dependence of the NN interaction in a 
model-dependent fashion. Indeed, with the recent developments in the 
production of polarized proton beams and the construction of high resolution 
spectrometers with focal plane polarimeters, it is possible to measure 
complete sets of polarization transfer observables which relate the 
components of a scattered polarized proton beam to the corresponding 
components of an incident proton beam which is polarized in an arbitrary 
direction [see Sec.~(\ref{sec:spinobs})].

To date most exclusive proton knockout data have been analyzed within 
the framework of the distorted wave impulse approximation (DWIA), 
the main ingredients of which are the scattering wave functions
for the incoming and two outgoing protons, the  boundstate wave function 
of the struck proton in the target nucleus, and the interaction between 
the incident proton and bound proton. Furthermore, the impulse 
approximation assumes that the form of the NN scattering matrix 
in the nuclear medium is the same as that for free NN scattering. 
The DWIA also assumes that the main influence of the nuclear medium 
is to modify (distort) the scattering wave functions relative to 
their corresponding plane wave values for scattering in free space: 
nuclear distortion effects are incorporated via the inclusion of 
appropriate optical potentials, gauged by elastic scattering data, 
in the underlying equations of motion. 

Since the tremendous success of the relativistic mean-field theory \cite {Se86} 
for describing nuclear reactions and nuclear structure, there are serious concerns 
regarding the validity of nonrelativistic Schr\"{o}dinger-equation-based models 
in nuclear physics. In this paper, we focus on a relativistic description of 
exclusive ($\vec{p},2 \vec{p}\,$) spin observables. Conventional wisdom claims 
that, since the binding energy of a nucleon in a nucleus is relatively small 
compared to the rest mass of a nucleon, relativistic effects are unimportant 
for nuclear structure problems, and hence the nonrelativistic Schr\"{o}dinger 
equation should provide an appropriate dynamical basis for nuclear physics 
studies. In recent years, however, the ability of quantum hadrodynamics, 
an effective relativistic field theory, to provide a mechanism for nuclear 
saturation and spin-orbit splitting in nuclei, has led to growing evidence 
that the relativistic Dirac equation is the correct underlying dynamical 
equation. In particular, the small nuclear binding energy and the strength 
of the spin-orbit interaction both result from the subtle interplay between 
an attractive Lorentz scalar (attributed to the exchange of sigma mesons), 
with a strength of approximately $-$400 MeV, and a repulsive vector potential 
(attributed to the exchange of omega mesons) with a strength of approximately 
$+$350 MeV \cite{Se86}. 

Recently, we demonstrated that the relativistic DWIA provides 
an excellent description of analyzing power data for the knockout of 
protons from the 3s$_{1/2}$, 2d$_{3/2}$ and  2d$_{3/2}$ and states in 
$^{208}$Pb at an incident energy of 202 MeV and for coincident coplanar 
scattering angles ($28.0^{\circ}$,$-54.6^{\circ}$)  \cite{Hi03}. Our motivation 
for choosing the $^{208}$Pb target and a relatively low incident energy of 
202 MeV was to maximize the influence of distortion effects, while still 
maintaining the validity of the impulse approximation, and also avoiding 
complications associated with the inclusion of recoil corrections in 
the relativistic Dirac equation \cite{Co93a,Ma93a}. In particular, we 
studied the effect of medium-modified coupling constants 
and meson masses on the above analyzing powers for both zero-range (ZR) and 
finite-range (FR) approximations to the relativistic DWIA. On one hand, 
the relativistic ZR predictions suggested that the scattering matrix for 
NN scattering in the nuclear medium is adequately represented by the 
corresponding matrix for free NN scattering, without nuclear medium corrections. On the 
other hand, the relativistic FR results imply that a 10\% to 20\% reduction 
of meson-coupling constants and meson masses by the nuclear medium is essential 
for providing a consistent description of the 3s$_{1/2}$, 2d$_{3/2}$ and 
2d$_{5/2}$ analyzing powers. Hence, within the context of the relativistic 
DWIA, it is not clear whether nuclear-medium modifications are important or not.
In addition, one needs to fully understand whether the differences between ZR and 
FR calculations are attributed to essential physics or numerical errors due to 
extensive computational procedures associated with FR predictions (compared to 
ZR calculations). In Refs.~\cite{Ne02,Hi03} it was also reported that the 
nonrelativistic Schr\"{o}dinger-equation DWIA predictions completely fail 
to reproduce the 3s$_{1/2}$ and 2d$_{3/2}$ analyzing powers. Systematic 
corrections to the nonrelativistic model - such as different kinematic 
prescriptions for the NN amplitudes, non-local corrections 
to the scattering wave functions, density-dependent modifications to the free 
NN scattering amplitudes, as well as the influence of different scattering and 
boundstate potentials - failed to remedy the nonrelativistic 
dilemma \cite{Ne02}. Although the analyzing power results seem to suggest 
that the Dirac equation is the preferred dynamical equation, a more definite 
statement regarding the role of dynamics can only be made after comparing 
model predictions to complete sets of polarization transfer observables 
for proton knockout from a variety of states in nuclei. In addition, such
a comparison will deepen our understanding of the influence of the nuclear 
medium effects on the NN interaction as well as shed light on the role of 
FR versus ZR effects in exclusive proton knockout reactions.

Unfortunately, there are no published spin observable data, other than the analyzing 
power, for the reaction kinematics of interest. In an effort to demonstrate 
the unique ability of polarization data, and in particular data on complete 
sets of polarization transfer observables, to selectively address many of
the above-mentioned physics issues, we present the first relativistic and 
nonrelativistic predictions of complete sets of polarization transfer observables 
for exclusive proton knockout from the 3s$_{1/2}$, 2d$_{3/2}$ and 2d$_{5/2}$ 
states in $^{208}$Pb, at an incident laboratory kinetic energy of 202 MeV, and 
for coincident coplanar scattering angles ($28.0^{\circ}$, $-54.6^{\circ}$). More 
specifically we investigate the sensitivity of these observables to FR versus ZR 
approximations to the relativistic DWIA as well as medium-modified meson-nucleon 
coupling constants and meson masses. In order to reliably extract
information on the latter it is necessary to minimize model-input uncertainties. 
The most likely source of uncertainty could be related to ambiguities associated 
with the choice of global optical potential parameters for generating
the incident and outgoing scattering wave functions: different global parameter sets are 
constrained by different sets of experimental data for elastic proton-nucleus 
scattering. For a heavy target nucleus such as $^{208}$Pb the effect of nuclear 
distortion is to reduce the unpolarized triple differential cross section to 
about 5\% of its plane wave value: differences in optical potential parameter
sets translate to an uncertainty of 10\% in the latter cross sections 
\cite{Co95}. The question arises as to how sensitive polarization transfer observables 
are to nuclear distortion and, in particular, to different optical potential parameter 
sets. Current qualitative arguments suggest that, since polarization transfer observables
are ratios of polarized cross sections, distortion effects on the scattering wave 
functions effectively cancel, and hence simple plane wave models (ignoring nuclear 
distortion) should be appropriate for studying polarization phenomena \cite{Ho88,Ho94}. 
Recently we demonstrated that, contrary to intuition, the ($\vec{p},2p$) analyzing 
power is extremely sensitive to nuclear distortion within the context of the relativistic 
DWIA \cite{Hi03}.  The analyzing power is, however, relatively insensitive to different 
global Dirac optical potential parameter sets. In this paper we extend the latter 
investigation to study, for the first time, the effect of nuclear distortion on 
complete sets of polarization transfer observables for exclusive ($\vec{p},2\vec{p}\,$) 
reactions. 

In Sec.~(\ref{sec:rdwia}), we briefly describe the essential ingredients underlying 
the relativistic DWIA for both ZR and FR approximations to the NN interaction. Thereafter, 
in Sec.~(\ref{sec:nucmed}), we discuss our prescription for invoking nuclear medium 
modifications of the NN interaction. The formalism for calculating complete sets of 
spin observables is presented in Sec.~(\ref{sec:spinobs}). Results are presented in
Sec.~(\ref{sec:results}), and we summarize and draw conclusions in 
Sec.~(\ref{sec:summary-and-conclusions}).

\section{\label{sec:rdwia}
Relativistic Distorted Wave Impulse Approximation}
Both ZR and FR approximations to the relativistic DWIA have been
discussed in detail in Refs. \cite{Ik95} and \cite{Ma96,Ma98}, respectively.
In this section, we briefly describe the main ingredients of these
models. The exclusive ($p,2p$) reaction of interest is schematically 
depicted in Fig.~(\ref{fig-p2pgeometry}), whereby an incident 
proton, $a$, knocks out a bound proton, $b$,
from a specific orbital in the target nucleus $A$, resulting in 
three particles in the final state, namely the recoil residual 
nucleus, $C$, and two outgoing protons, $a'$ and $b$, which are 
detected in coincidence at coplanar laboratory scattering angles, 
$\theta_{a'}$ and $\theta_{b}$, respectively. All kinematic
quantities are completely determined by specifying 
the rest masses, $m_{i}$, of particles, where 
$i$ = ($a$,$A$, $a'$, $b$, $C$), the laboratory kinetic energy, 
$T_{a}$, of incident particle $a$, the laboratory kinetic energy, 
$T_{a'}$, of scattered particle $a'$, the laboratory scattering angles 
$\theta_{a'}$ and $\theta_{b'}$, and also the binding energy 
of the proton that is to be knocked out of the target nucleus, $A$.

\begin{figure}
\includegraphics{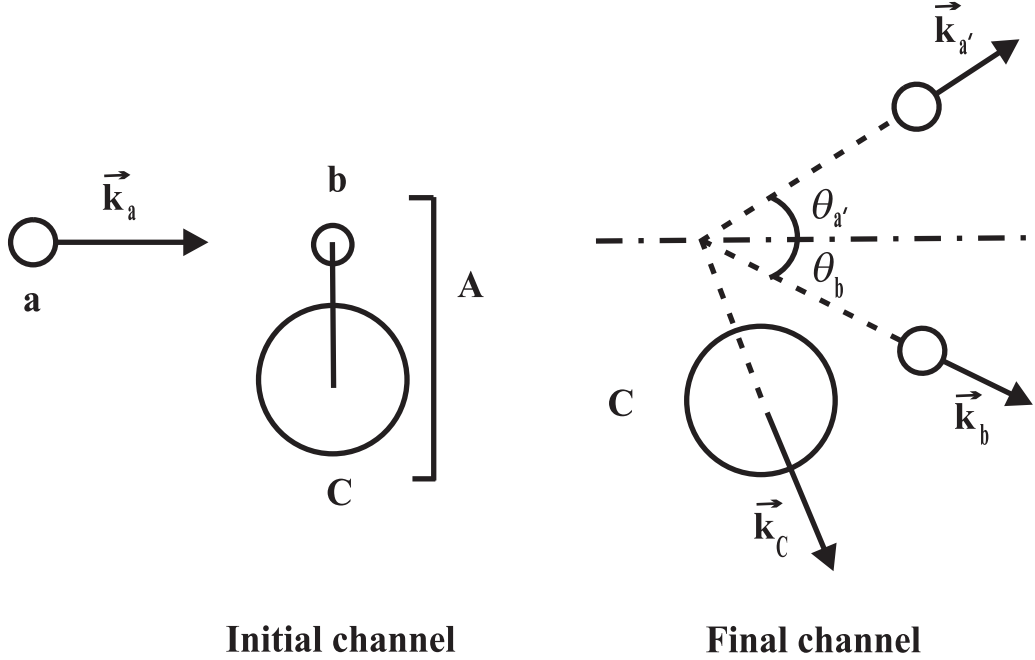}
\caption{\label{fig-p2pgeometry}Schematic representation for the 
coplanar $(p,2p)$ reaction of interest.}
\label{fig-p2pgeometry }
\end{figure}

For a finite-range (FR) NN interaction, the relativistic distorted wave 
transition matrix element is given by
\begin{eqnarray}
\hspace{-7mm} T_{L J M_J}(s_{a}, s_{a'}, s_{b}) & = & 
\int d\vec{r}\ d\vec{r}\,'\  
[
\bar{\psi}^{(-)}(\vec{r}, \vec{k}_{a'C},s_{a'}) 
\otimes   
\bar{\psi}^{(-)}(\vec{r}\,', \vec{k}_{bC},s_{b})\,
]\,  \times \nonumber\\
 & & \ \ \ \hat{t}_{NN}(\, | \vec{r} - \vec{r}\,'| \, )
[
\psi^{(+)}(\vec{r}, \vec{k}_{aA}, s_{a})
\otimes
\phi^{B}_{L J M_J}(\vec{r}\,'\,)
]
\label{e-frtjlm}
\end{eqnarray}
where $\otimes$ denotes the Kronecker product. The four-component
scattering wave functions, $\psi(\vec{r}, \vec{k}_{ij},s_{i})$, 
are solutions to the fixed-energy Dirac equation with spherical scalar, 
$S(r)$, and time-like vector, $V(r)$, nuclear optical potentials: 
$\psi^{(+)}(\vec{r}, \vec{k}_{aA},s_{a})$ is the relativistic scattering 
wave function of the incident particle, $a$, with outgoing boundary 
conditions [indicated by the superscript $(+)$], where $\vec{k}_{aA}$ 
is the momentum of particle $a$ in the ($a$ + $A$) center-of-mass system, 
and $s_{a}$ is the spin projection of particle $a$ with respect to 
$\vec{k}_{aA}$ as the $\hat{z}$-quantization axis; 
$\bar{\psi}^{(-)}(\vec{r}, \vec{k}_{j C},s_{j})$
is the adjoint relativistic scattering wave function for particle $j$ 
[ $j$ = ($a',b$)] with incoming boundary conditions [indicated by the 
superscript $(-)$], where $\vec{k}_{j C}$ is the momentum of particle
$j$ in the ($j$ + $C$) center-of-mass system, and $s_{j}$ is the spin projection 
of particle $j$ with respect to $\vec{k}_{j C}$ as the $\hat{z}$-quantization axis.
The boundstate proton wave function, $\phi^{B}_{L J M_J}(\vec{r}\, )$, 
labeled by single-particle quantum numbers $L$, $J$, and $M_{J}$, is obtained 
via selfconsistent solution to the Dirac-Hartree field equations of 
quantum hadrodynamics \cite{Ho81}. In addition, we adopt the impulse
approximation which assumes that the form of the NN scattering matrix
in the nuclear medium is the same as that for free NN scattering.
Furthermore, we assume that the antisymmetrized NN scattering matrix, 
$\hat{t}_{NN}(\, | \vec{r} - \vec{r}\,'| \, )$, 
is parameterized in terms of the five Fermi covariants \cite{Mc83a}, the
so-called IA1 representation of the NN scattering amplitudes. 
In principle, the NN $t$-matrix can be 
obtained via solution of the Bethe-Salpeter equation, where the on-shell 
NN amplitudes are matrix elements of this $t$-matrix. However, the complexity 
of this approach gives limited physical insight into the resulting amplitudes.
An alternative approach is to fit the amplitudes directly with some 
phenomenological form, rather than generating the $t$-matrix from a 
microscopic interaction. Although the microscopic approach is certainly 
more fundamental, the advantage of phenomenological fits lies in their 
simple analytical form, which allows them to be conveniently incorporated 
in calculations requiring the NN $t$-matrix as input. The NN $t$-matrix 
employed in this paper is based on the relativistic meson-exchange model 
described in Ref.~\cite{Ho85}, the so-called relativistic Horowitz-Love-Franey 
(HLF) model, where the direct and exchange contributions to the IA1 amplitudes 
are parameterized separately in terms of a number of Yukawa-type meson exchanges 
in first-order Born approximation. The parameters of this interaction, namely 
the meson masses, meson-nucleon coupling constants and the cutoff parameters, 
have been adjusted to reproduce the free NN elastic scattering observables.

Adopting a much simpler zero-range (ZR) approximation for the NN interaction, namely
\begin{eqnarray}
\hat{t}_{NN}(\, | \vec{r} - \vec{r}\,'| \, )\ =\ 
\hat{t}_{NN}(T_{\rm eff}^{\rm \ell ab}, \theta_{\rm eff}^{\rm cm})
\, \delta(\vec{r} - \vec{r}\,'\, )
\label{e-zra}
\end{eqnarray} 
the relativistic distorted wave transition matrix element in Eq.~(\ref{e-frtjlm})
reduces to
\begin{eqnarray}
\hspace{-7mm} T_{L J M_J}(s_{a}, s_{a'}, s_{b}) & = & 
\int d\vec{r}\, 
[\,
\bar{\psi}^{(-)}(\vec{r}, \vec{k}_{a'C},s_{a'}) 
\otimes   
\bar{\psi}^{(-)}(\vec{r}, \vec{k}_{bC},s_{b})
\,] \nonumber\\
 & & \hat{t}_{NN}(T_{\rm eff}^{\rm \ell ab}, \theta_{\rm eff}^{\rm cm})\,
[\,
\psi^{(+)}(\vec{r}, \vec{k}_{aA}, s_{a})
\otimes
\phi^{B}_{L J M_J}(\vec{r}\,)
\,]
\label{e-tjlm}
\end{eqnarray}
where $T_{\rm eff}^{\rm \ell ab}$ and $\theta_{\rm eff}^{\rm cm}$ 
represent the effective two-body laboratory kinetic energy and 
center-of-mass scattering angles, respectively. 

As already mentioned, a FR approximation to the DWIA is inherently more 
sophisticated than a ZR approximation. However, in practice, the numerical 
evaluation of the six-dimensional FR transition matrix elements, 
given by Eq.~(\ref{e-frtjlm}), is nontrivial and subject 
to numerical uncertainties. On the other hand, for the ZR approximation, 
the three-dimensional integral given by Eq.~(\ref{e-tjlm}), 
ensures numerical stability and rapid convergence (and hence faster 
computational time). Another  advantage of the ZR approximation is that 
one can directly employ experimental NN scattering amplitudes, rather than rely on a 
relativistic meson-exchange model, and hence, one is insensitive to 
uncertainties associated with interpolations and/or extrapolations of the limited 
meson-exchange parameter sets. In this paper, we compare FR and ZR 
predictions for complete sets of polarization transfer observables.

In principle, one could employ the HLF model for also generating microscopic
relativistic scalar and vector optical potentials by folding the NN $t$-matrix
with the appropriate Lorentz densities via the $t \rho$ approximation. An 
attractive feature of the $t \rho$ approximation is selfconsistency, that is, 
the HLF model is used for generating both NN scattering amplitudes and optical 
potentials. However, for the kinematic region of interest to this paper, 
we consider it inappropriate to employ microscopic $t \rho$ optical potentials, 
the reason being that HLF parameter sets only exist at 135 MeV and 200 MeV, 
whereas optical potentials for the outgoing protons are required at energies 
ranging between 24 and 170 MeV. Thus, enforcing selfconsistency would involve 
large, and relatively crude, interpolations/extrapolations, leading to inaccurate
predictions of the spin observables. Furthermore, the validity of the impulse 
approximation, to generate microscopic $t \rho$ optical potentials at energies 
lower than 100 MeV, is questionable. Hence, in this paper we consider only 
global Dirac optical potentials \cite{Co93b}, as opposed to microscopic $t \rho$ 
optical potentials, for obtaining the scattering wavefunctions of the Dirac equation. 
\section{\label{sec:nucmed}Nuclear medium effects} 
For estimating the influence of nuclear medium modifications of the NN 
interaction on spin observables, we adopt the Brown-Rho scaling conjecture \cite{Br91} 
which attributes nuclear-medium modifications of meson-nucleon coupling constants, 
as well as nucleon- and meson-masses, to partial restoration of chiral symmetry.
In particular, we invoke the scaling relations proposed by Brown and Rho 
\cite{Br91}, and also applied by Krein {\it et al.} \cite{Kr95}
to ($p,2p$) reactions, namely
\begin{eqnarray}
\frac{m_{\sigma}^{\ast}}{m_{\sigma}}
& \approx &  
\frac{m_{\rho}^{\ast}}{m_{\rho}} \approx \frac{m_{\omega}^{\ast}}
{m_{\omega}}\equiv \xi \label{eqn:a2}\ ,\\
\frac{g_{\sigma N}^{\ast}}{g_{\sigma N}}
 & \approx &
\frac{g_{\omega N}^{\ast}}{g_{\omega N}} \equiv \chi \label{eqn:a3}\ ,
\end{eqnarray}
where the medium-modified and free meson masses
are denoted by $m_{i}^{\ast}$ and $m_{i}$, with
$i \in \mbox{(}\sigma,\rho,\omega\mbox{)}$, respectively. Meson-nucleon 
coupling constants, with and without nuclear medium modifications, are 
denoted by $g_{j N}^{\ast}$ and $g_{j N}$, where $j \in \mbox{(}\sigma,\omega\mbox{)}$, 
respectively: see Sec.~(\ref{sec:results}) for typical values of $\xi$ and $\chi$.
\section{\label{sec:spinobs}Spin observables}
The spin observables of interest are denoted by $\mbox{D}_{i' j}$ and are 
related to the probability that an incident beam of particles, $a$, 
with spin-polarization $j$ induces a spin-polarization $i'$ for 
the scattered beam of particles, $a'$: the subscript $j = (0,\ell,n,s)$ 
is used to specify the polarization of the incident beam, $a$, along any 
of the orthogonal directions
\begin{eqnarray}
\hat{\ell}\ =\ \hat{z}\ =\ \hat{k}_{aA} \nonumber\\
\hat{n}\ =\ \hat{y}\ =\ \hat{k}_{aA} \times \hat{k}_{a' C} \nonumber\\
\hat{s}\ =\ \hat{x}\ =\ \hat{n} \times \hat{\ell}\,,
\label{e-lns}
\end{eqnarray}
and the subscript $i' = (0,\ell',n',s')$ denotes the polarization of the scattered 
beam, $a'$, along any of the orthogonal directions:
\begin{eqnarray}
\hat{\ell}'\ =\ \hat{z}'\ =\ \hat{k}_{a' C} \nonumber\\
\hat{n}'\ =\  \hat{n}\ =\ \hat{y} \nonumber\\
\hat{s}'\ =\ \hat{x}'\ =\ \hat{n} \times \hat{\ell}'\,.
\label{e-lpnsp}
\end{eqnarray}
The choice $j\,(i') = 0$ is used to denote an unpolarized incident (scattered) beam.
With the above coordinate axes in the initial and final channels, the 
spin observables, $\mbox{D}_{i' j}$, are defined by
\begin{eqnarray}
\mbox{D}_{i' j}\ =\ \frac{\mbox{Tr}(T \sigma_{j} T^{\dagger} \sigma_{i'})}
{\mbox{Tr}(T T^{\dagger})}\,,
\label{e-dipj}
\end{eqnarray}
where $\mbox{D}_{n 0}=\mbox{P}$ refers to the induced polarization, 
$\mbox{D}_{0 n}=\mbox{A}_{y}$ denotes the analyzing power, and the other
polarization transfer observables of interest are
$\mbox{D}_{n n},\, \mbox{D}_{s' s},\, \mbox{D}_{\ell' \ell},\, 
\mbox{D}_{s' \ell},\ \mbox{and}\ \mbox{D}_{\ell' s}$. 
The denominator of Eq.~(\ref{e-dipj}) is related to the unpolarized triple 
differential cross section, i.e.,
\begin{eqnarray}
\frac{d^3 \sigma}
{d T_{a'}\, d \Omega_{a'}\, d \Omega_b}\ \ \ \propto\ \ \    
\mbox{Tr}(T\, T^{\dagger})\,.
\label{e-unpol}
\end{eqnarray}
In Eq.~(\ref{e-dipj}), the symbols $\sigma_{i'}$ and $\sigma_{j}$ 
denote the usual $2 \times 2$ Pauli spin matrices, namely,
\begin{eqnarray}
\sigma_{0}\ =\ 
\left(\begin{array}{cc}
 1 & 0 \\
 0 & 1 
\end{array}\right)\nonumber\\
\sigma_{s'}\ =\ \sigma_{s}\ =\ \sigma_{x}\ =\ 
\left(\begin{array}{cc}
 0 & 1 \\
 1 & 0 
\end{array}\right)\nonumber\\
\sigma_{n}\ =\ \sigma_{y}\ =\  
\left(\begin{array}{cc}
 0 & -i \\
 i & 0 
\end{array}\right)\nonumber\\
\sigma_{\ell'}\ =\ \sigma_{\ell}\ =\ \sigma_{z}\ =\ 
\left(\begin{array}{cc}
 1 & 0 \\
 0 & -1 
\end{array}\right)
\label{e-paulism}
\end{eqnarray}
and the $2 \times 2$ matrix $T$ is given by
\begin{eqnarray}
T\ =\ \left(
\begin{array}{cc}
T_{L J}^{s_{a} = +\frac{1}{2}, s_{a'} = +\frac{1}{2}} & 
T_{L J}^{s_{a} = -\frac{1}{2}, s_{a'} = +\frac{1}{2}} \\
T_{L J}^{s_{a} = +\frac{1}{2}, s_{a'} = -\frac{1}{2}} & 
T_{L J}^{s_{a} = -\frac{1}{2}, s_{a'} = -\frac{1}{2}} 
\end{array}\right)
\label{e-ttwobytwo}
\end{eqnarray}
where $s_{a} = \pm \frac{1}{2}$ and $s_{a'} = \pm \frac{1}{2}$ refer to the spin
projections of particles $a$ and $a'$ along the $\hat{z}$ and $\hat{z}'$ axes,
defined in Eqs.~(\ref{e-lns}) and (\ref{e-lpnsp}), respectively;
the matrix $T_{L J}^{s_{a}, s_{a'}}$ is related to the relativistic $(p,2p)$ transition
matrix element $T_{L J M_{J}}(s_{a}, s_{a'}, s_{b})$, defined in 
Eqs.~(\ref{e-frtjlm}) and (\ref{e-tjlm}), via
\begin{eqnarray}
T_{L J}^{s_{a}, s_{a'}}\ =\ \sum_{M_J, s_b} T_{L J M_J}
(s_a, s_{a'}, s_b)\,.
\label{e-tljmrelatet}
\end{eqnarray}

\section{\label{sec:results}Results}
In this section we study the sensitivity of complete sets
of exclusive ($\vec{p},2 \vec{p}\,$) spin observables, 
for the knockout of protons from the 3s$_{1/2}$, 2d$_{3/2}$ 
and 2d$_{5/2}$ states in $^{208}$Pb, at an incident energy of 
202 MeV, and for coincident coplanar scattering angles 
($28.0^{\circ}$,$-54.6^{\circ}$), to distorting optical 
potentials, FR versus ZR approximations to the relativistic 
DWIA, as well as to medium-modified meson-nucleon coupling constants and 
meson masses. We also compare our relativistic results to 
corresponding nonrelativistic Schr\"{o}dinger-based predictions 
based on the computer code THREEDEE of Chant and Roos \cite{Ch}. 
Our aim is to identify specific observables which can be measured
in order to unravel and understand the role of the above approximations,
model ingredients and different dynamical models. Unless otherwise 
specified, all DWIA predictions employ the energy-dependent global 
Dirac optical potential parameter set constrained by $^{208}$Pb($p,p$) 
elastic scattering data for incident proton energies between 21 MeV and 
1040 MeV \cite{Co93b}.

\begin{figure}[htb]
\includegraphics{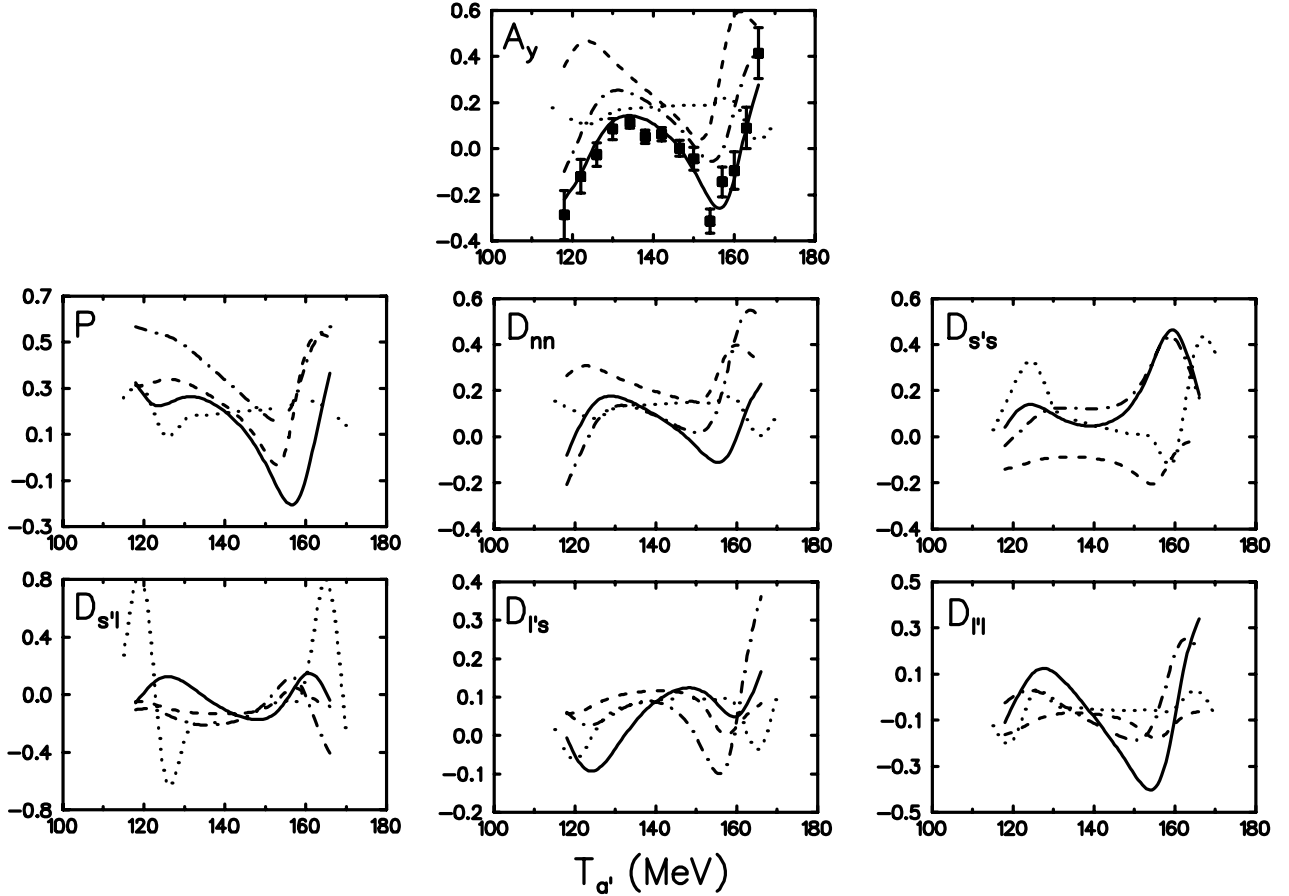}
\caption{\label{fig:fig2} 
Polarization transfer observables plotted as a function of the kinetic 
energy, $T_{a'}$, for the knockout of protons from the 3s$_{1/2}$ state 
in $^{208}$Pb, at an incident energy of 202 MeV, and for coincident coplanar 
scattering angles ($28.0^{\circ}$, $-54.6^{\circ}$). The different line types 
represent the following calculations: relativistic ZR-DWIA (solid line), relativistic 
plane wave (dotted line), nonrelativistic DWIA (dashed line), and relativistic 
FR-DWIA (dot-dashed line). The analyzing power data are from Ref.~\cite{Ne02}.}
\end{figure}

\begin{figure}[htb]
\includegraphics{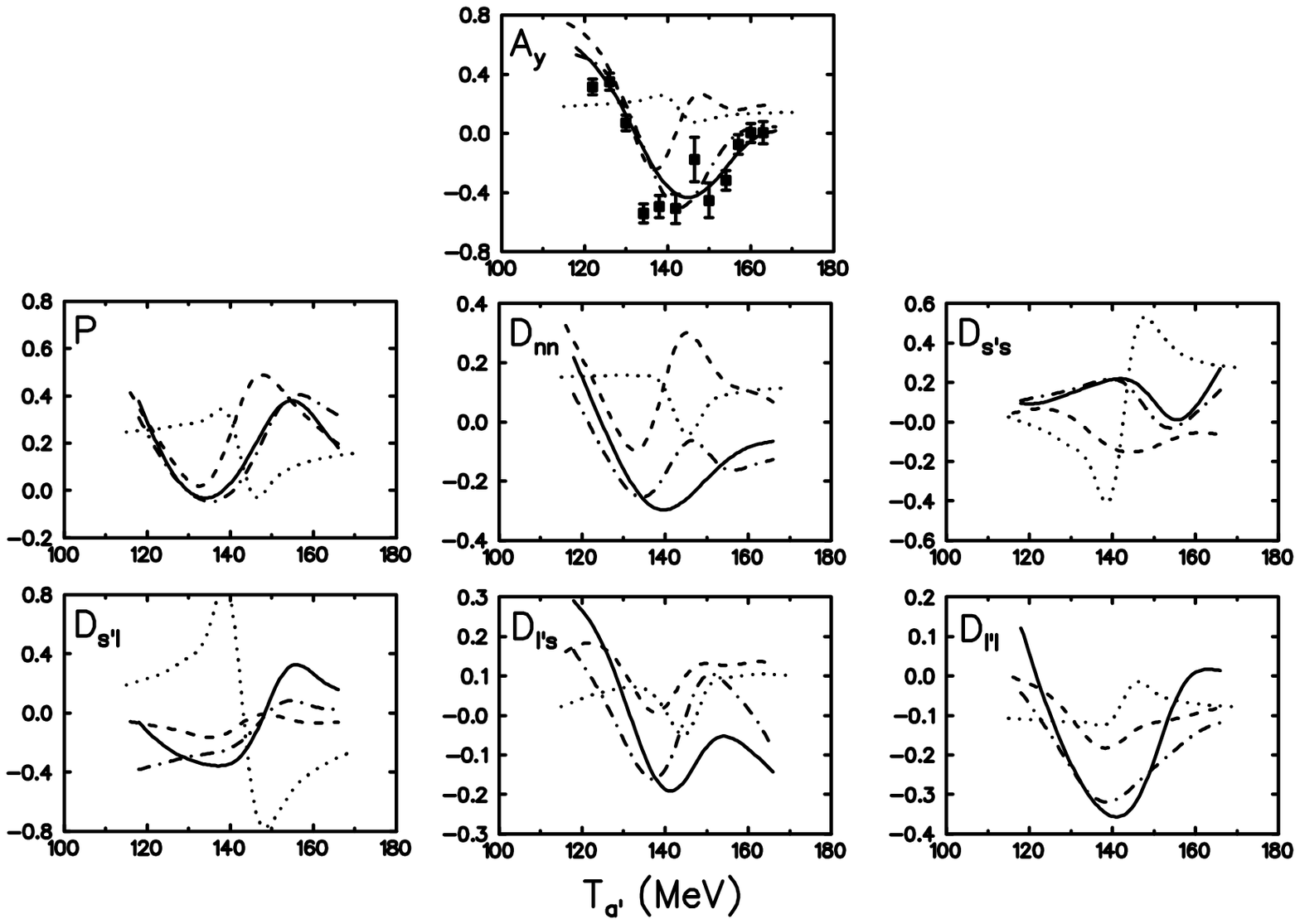}
\caption{\label{fig:fig3} 
Polarization transfer observables plotted as a function of the kinetic 
energy, $T_{a'}$, for the knockout of protons from 
the 2d$_{3/2}$ state in $^{208}$Pb, at an incident energy of 202 MeV, and 
for coincident coplanar scattering angles ($28.0^{\circ}$, $-54.6^{\circ}$).
The different line types represent the following calculations:
relativistic ZR-DWIA (solid line), relativistic plane wave (dotted
line), nonrelativistic DWIA (dashed line), and relativistic 
FR-DWIA (dot-dashed line). The analyzing power data are from Ref.~\cite{Ne02}.}
\end{figure}

\begin{figure}[htb]
\includegraphics{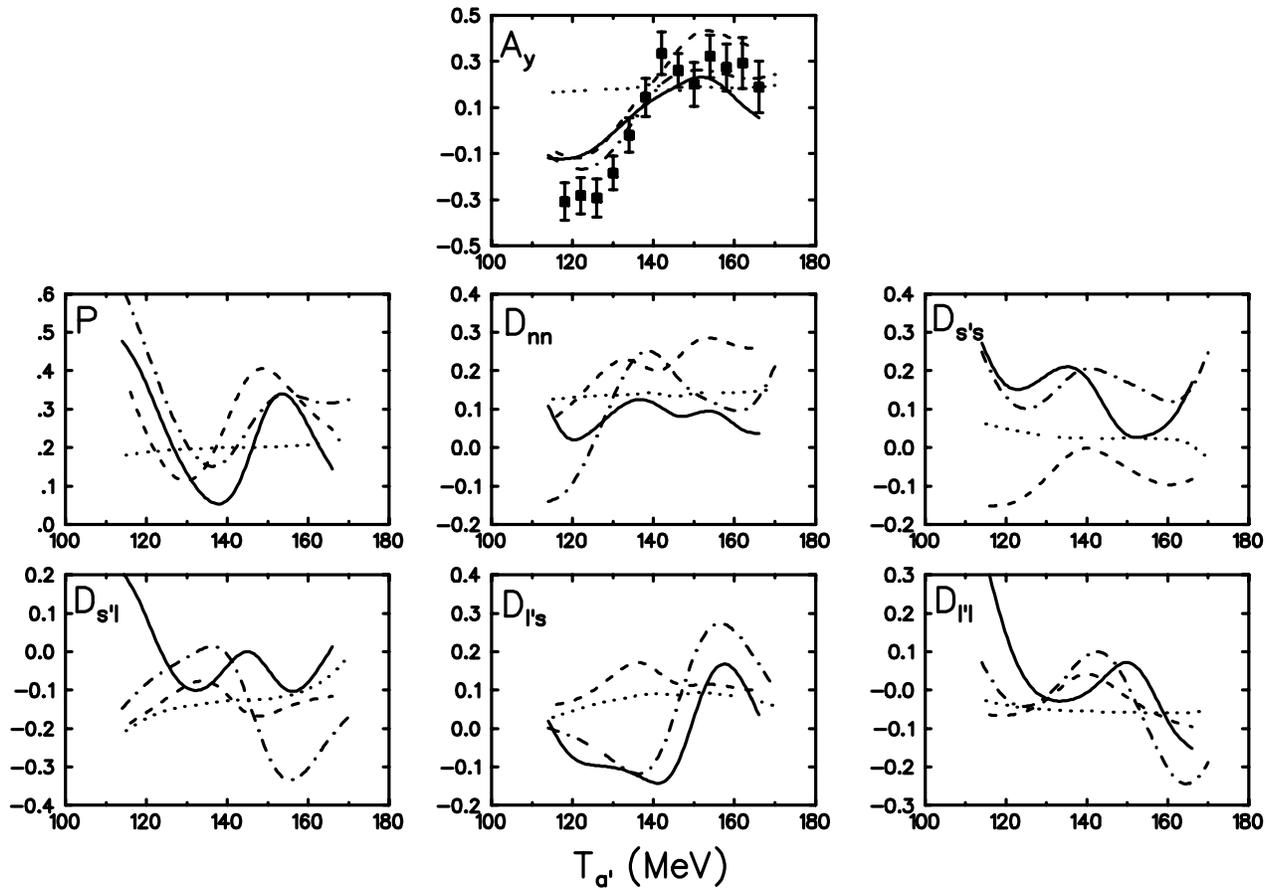}
\caption{\label{fig:fig4} 
Polarization transfer observables plotted as a function of the kinetic 
energy, $T_{a'}$, for the knockout of protons from the 2d$_{5/2}$ state 
in $^{208}$Pb, at an incident energy of 202 MeV, and for coincident 
coplanar scattering angles ($28.0^{\circ}$, $-54.6^{\circ}$).
The different line types represent the following calculations:
relativistic ZR-DWIA (solid line), relativistic plane wave (dotted
line), nonrelativistic DWIA (dashed line), and relativistic 
FR-DWIA (dot-dashed line). The analyzing power data are from Ref.~\cite{Ne02}.}
\end{figure}

First, we display the influence of relativistic nuclear distortion 
effects on complete sets of spin observables by comparing relativistic 
ZR-DWIA to relativistic plane wave predictions (with zero scattering potentials) 
for knockout from the 3s$_{1/2}$, 2d$_{3/2}$ and 2d$_{5/2}$ states 
in Figs.~(\ref{fig:fig2}), (\ref{fig:fig3}) and (\ref{fig:fig4}), respectively: 
the solid lines indicate the relativistic ZR distorted wave result and the 
dotted lines represent the relativistic plane wave result.
For completeness we include the analyzing power calculations 
reported in Ref.~\cite{Hi03}. As already mentioned in the
latter publication, we see that the prominent oscillatory 
structure of the analyzing powers is mostly attributed to 
distortions of the scattering wave functions. Similarly, for the other 
spin observables there are large differences between the 
relativistic distorted wave and plane wave results. The above observations 
clearly illustrate the importance of including nuclear distorting 
optical potentials for calculating spin observables, thus 
refuting previous qualitative claims that spin observables, 
being ratios of cross sections, are insensitive to nuclear 
distortion effects. In addition, we have also investigated 
the sensitivity of all spin observables to a variety of different 
global Dirac optical potential parameter sets \cite{Co93b}. 
Although these results are not displayed, we find that all spin 
observables are relatively insensitive to different global 
optical potentials, with differences between parameter sets 
being smaller than the experimental statistical error indicated 
on the analyzing powers.

Next, we compare relativistic FR-DWIA (dot-dashed line) to relativistic 
ZR-DWIA (solid line) predictions. In general, it is seen that most spin 
observables are relatively sensitive to differences in ZR and FR predictions. 
For knockout from the 3s$_{1/2}$ state [Fig.~(\ref{fig:fig2})], the induced 
polarization, P, is the most sensitive observable to differences between FR and ZR 
predictions, whereas D$_{s's}$ is relatively insensitive. For knockout from the 
2d$_{3/2}$ state [Fig.~(\ref{fig:fig3})], on the other hand, D$_{\ell' s}$ displays 
large differences between ZR and FR calculations, whereas A$_{y}$, P and D$_{s' s}$ 
display small differences. For the 2d$_{5/2}$ state [Fig.~(\ref{fig:fig4})], 
D$_{s' \ell}$ and A$_{y}$ are the most and least sensitive observables to FR versus 
ZR differences, respectively. When comparing ZR and FR predictions to the only 
existing proton knockout spin observable data on $^{208}$Pb, namely the analyzing 
power, we generally see that the ZR predictions provide an excellent description 
for knockout from all three states. Note, however, that although the relativistic FR 
(dot-dashed line) predictions are not as spectacular as the corresponding ZR calculations, 
they still provide a reasonable qualitative description of the data. The measurement
of observables which display large differences between ZR and FR predictions
will check the consistency of the analyzing power results and serve to further
constrain relativistic DWIA models.

We also compare our relativistic ZR and FR calculations to nonrelativistic [dashed 
line in  Figs.~(\ref{fig:fig2}), (\ref{fig:fig3}) and (\ref{fig:fig4})] 
DWIA predictions based on the commonly-used computer code THREEDEE of Chant 
and Roos \cite{Ch}. First, we mention the analyzing power results reported
in Ref.~\cite{Hi03}. With the exception of the 2d$_{5/2}$, it is clearly seen that 
the relativistic ZR (solid line) and FR (dot-dashed line) predictions
are consistently superior compared to the corresponding 
nonrelativistic calculations. This suggests that the Dirac equation is the 
most appropriate dynamical equation for the description of analyzing powers. 
Moreover, these results represent the clearest signatures to date for the 
evidence of relativistic dynamics in polarization phenomena. However, before 
claiming with certainty that the relativistic equation is the most appropriate 
dynamical equation, it is necessary to identify additional observables which 
should be measured in order to further study the question of dynamics. In this respect, 
we generally see that all spin observables are relatively sensitive to Dirac- 
versus Schr\"{o}dinger-based DWIA models. In particular, for knockout from the
3s$_{1/2}$ state the spin observables A$_{y}$, D$_{n n}$ and D$_{s' s}$ exhibit
large differences between Dirac and Schr\"{o}dinger-based DWIA models. On the
other hand, for knockout from the 2d$_{3/2}$ state the most sensitive observables
to dynamical differences are D$_{n n}$, D$_{\ell ' s}$ and D$_{\ell ' \ell}$.
For 2d$_{5/2}$ knockout the most sensitive observables are D$_{s' s}$ and 
D$_{\ell ' s}$. A number of interesting observations are made at the point
$T_{a'} \approx 145$ MeV corresponding to minimum recoil momentum.
First of all, we see that for 3s$_{1/2}$ knockout the induced polarization (P), 
D$_{s' \ell}$ and D$_{\ell ' s}$, the relativistic plane wave, ZR-DWIA and 
nonrelativistic DWIA predictions are virtually identical at this point; 
ZR-DWIA and FR-DWIA predictions are nearly identical for both D$_{n n}$ and 
D$_{s ' \ell}$. For knockout from the 2d$_{3/2}$ state, both FR-DWIA and ZR-DWIA
yield similar results for P, D$_{s' s}$ and D$_{s' \ell}$; P and 
D$_{s' s}$ are insensitive to nuclear distortion at the point in 
question. Finally, we see that for 2d$_{5/2}$ knockout, relativistic 
plane wave, FR-DWIA and nonrelativistic DWIA predictions are virtually 
identical for D$_{s' \ell}$ and D$_{\ell' s}$. Hence, by measuring 
spin observables at minimum recoil momentum one can eliminate differences 
between different dynamical models and model parameters and focus on a 
specific issue of interest.

\begin{figure}[htb]
\includegraphics{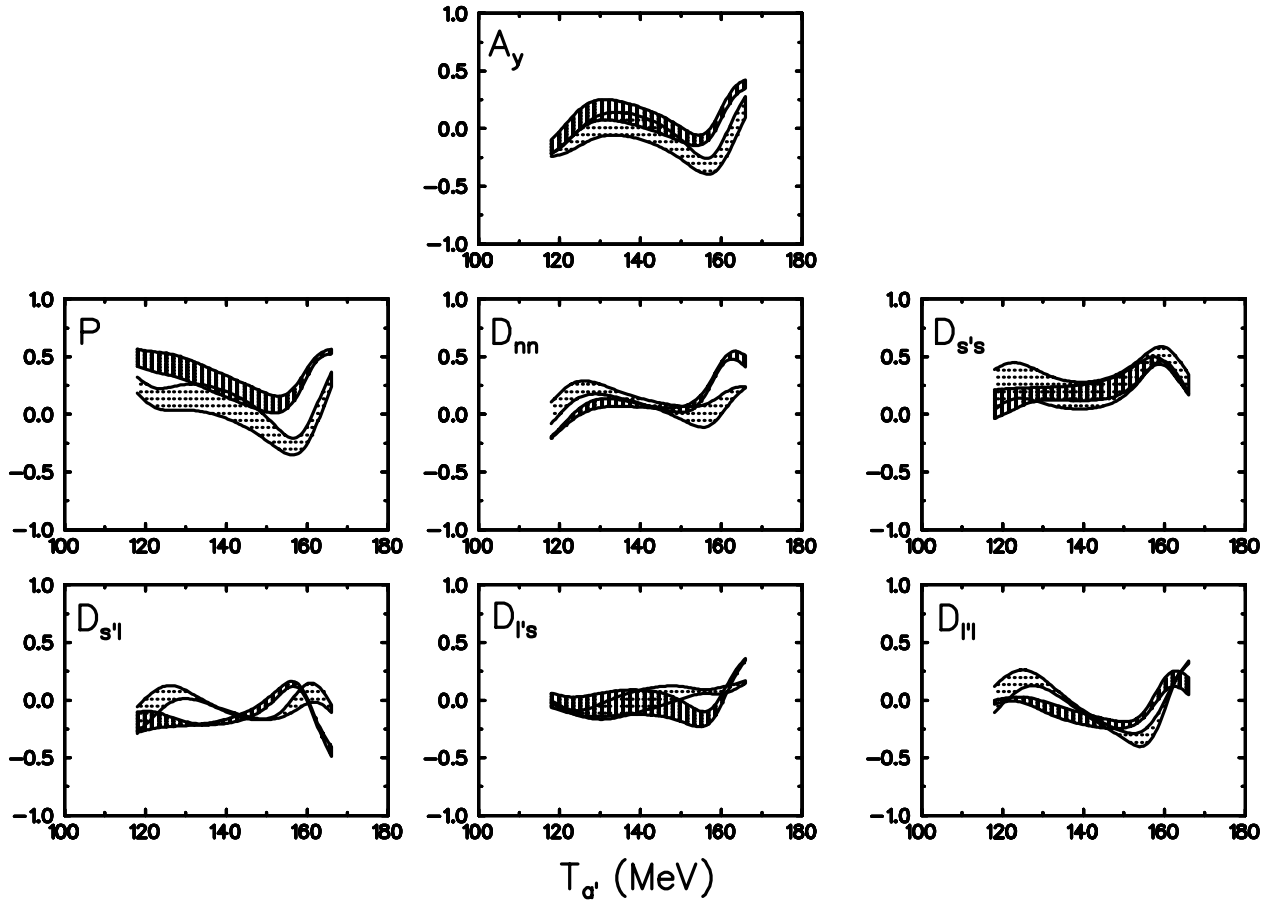}
\caption{\label{fig:fig5} 
Polarization transfer observables plotted as a function of the kinetic 
energy, $T_{a'}$, for the knockout of protons from the 3s$_{1/2}$ state 
in $^{208}$Pb, at an incident energy of 202 MeV, and for coincident 
coplanar scattering angles ($28.0^{\circ}$, $-54.6^{\circ}$). The vertically 
hatched band represents the sensitivity of a particular FR-DWIA spin 
observables to a reduction of coupling constants and meson masses ranging 
from 0\% to 20\% of the free values. The dotted band represents the 
corresponding ZR-DWIA predictions.}
\end{figure}

\begin{figure}[htb]
\includegraphics{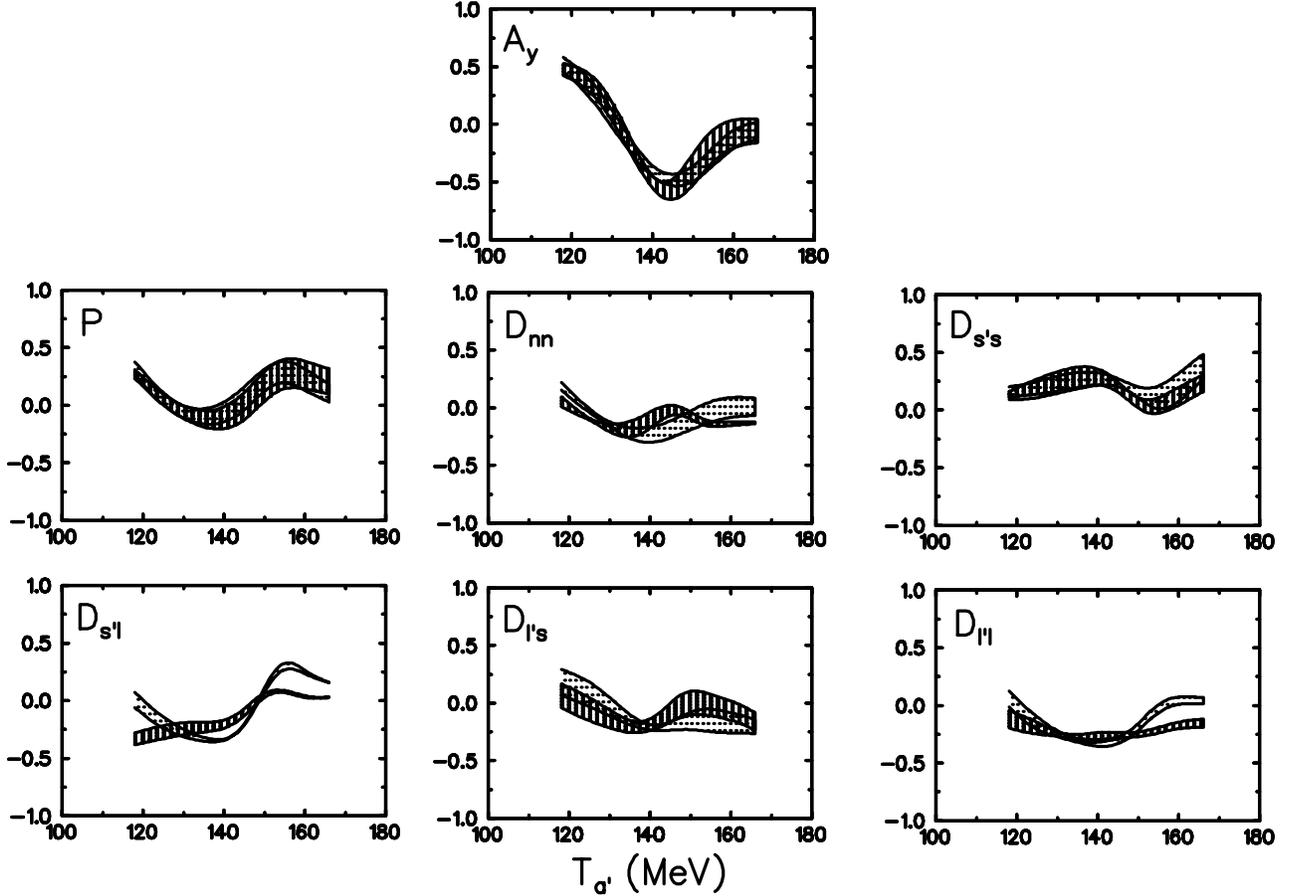}
\caption{\label{fig:fig6} 
Polarization transfer observables plotted as a function of the kinetic 
energy, $T_{a'}$, for the knockout of protons from the 2d$_{3/2}$ state 
in $^{208}$Pb, at an incident energy of 202 MeV, and for coincident 
coplanar scattering angles ($28.0^{\circ}$, $-54.6^{\circ}$). The vertically 
hatched band represents the sensitivity of a particular FR-DWIA spin 
observables to a reduction of meson-coupling constants and meson masses 
ranging from 0\% to 20\% of the free values. The dotted band represents 
the corresponding ZR-DWIA predictions.}
\end{figure}

\begin{figure}[htb]
\includegraphics{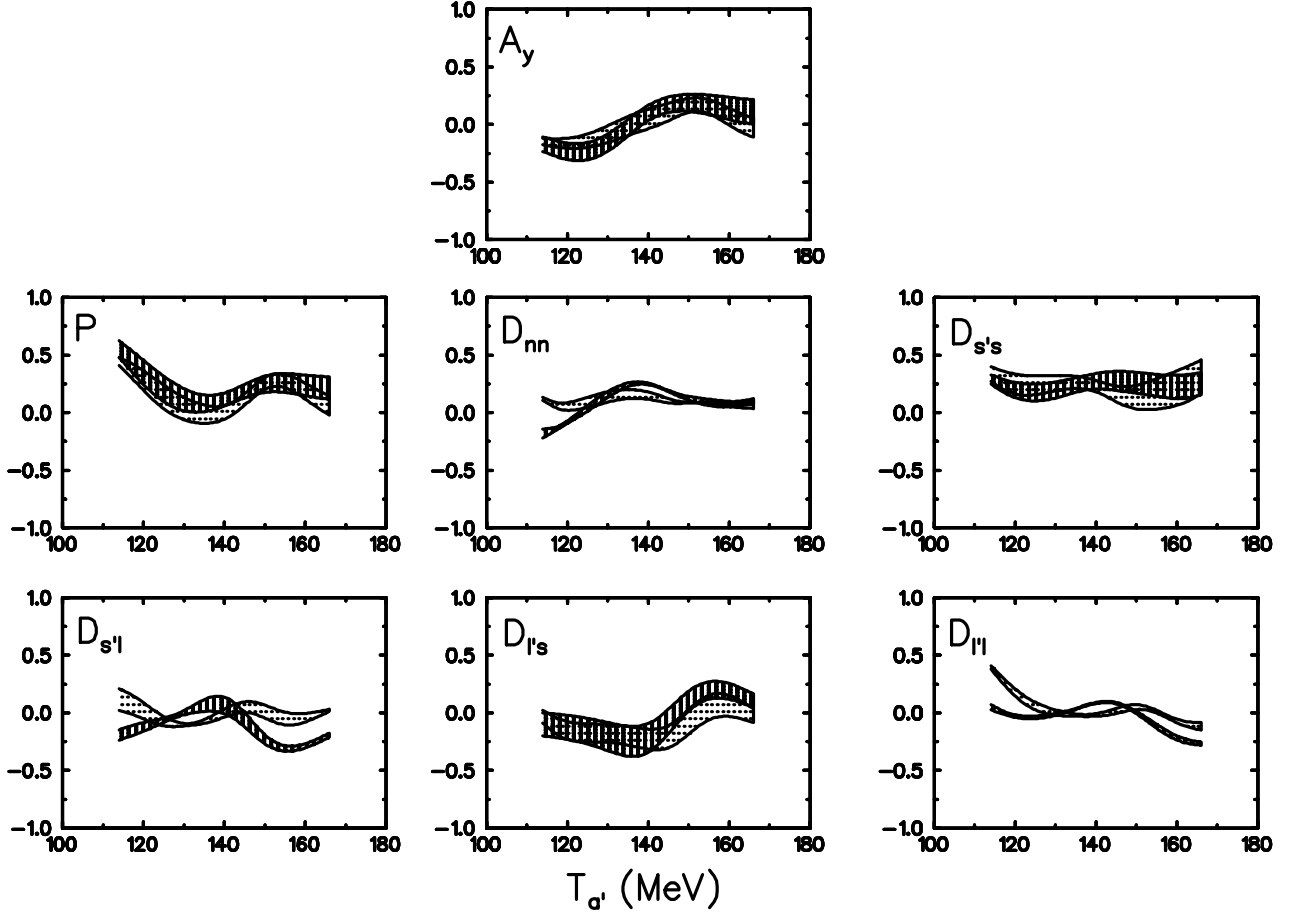}
\caption{\label{fig:fig7} 
Polarization transfer observables plotted as a function of the kinetic 
energy, $T_{a'}$, for the knockout of protons from the 2d$_{5/2}$ state 
in $^{208}$Pb, at an incident energy of 202 MeV, and for coincident 
coplanar scattering angles ($28.0^{\circ}$, $-54.6^{\circ}$). The vertically 
hatched band represents the sensitivity of a particular FR-DWIA spin 
observables to a reduction of meson-coupling constants and meson masses ranging 
from 0\% to 20\% of the free values. The dotted band represents the 
corresponding ZR-DWIA predictions.}
\end{figure}

Next we study the sensitivity of spin observables to the nuclear medium modifications 
of the NN interaction [discussed in Sec.~(\ref{sec:nucmed})] within the context
of the relativistic DWIA. In Ref.~\cite{Hi03} 
we studied the sensitivity of analyzing powers to 20\% reductions of meson-nucleon coupling 
constants and meson masses by the nuclear medium relative to the values for 
free NN scattering. More specifically we chose $\xi = \chi$ and varied these 
values between 1.0 and 0.8 for knockout from all three states of
interest. The latter equality is only assumed for simplicity, so as to get a feeling
for the sensitivity of observables to changes in the relevant meson-nucleon coupling 
constants and meson masses. The choice of values for $\xi$ and $\chi$ 
is motivated by the fact that the proton-knockout reactions of interest 
are mainly localized in the nuclear surface and, hence, the nuclear medium 
modifications are expected to play a relatively minor role. Actually, using the 
procedure proposed in Ref.~\cite{Ha97}, the effective mean densities are 
estimated to be between 0.08 and 0.15 of the saturation density. In particular, for the 
analyzing powers in question we established that for values of $\xi = \chi < 0.8$
both FR-DWIA and ZR-DWIA models fail to reproduce the experimental analyzing power data. 
Regarding nuclear medium effects, for the ZR predictions we concluded in Ref.~\cite{Hi03} that the 
inclusion of medium-modified meson-nucleon coupling constants and meson masses successfully 
described the analyzing power data, whereas the ZR predictions suggest that 
the scattering matrix for NN scattering in the nuclear medium is adequately 
represented by the corresponding matrix for free NN scattering, excluding 
corrections for the nuclear medium. It is important to measure other spin
observable data in order to check the consistency of the conclusion based on
only the analyzing power data. In this paper we investigate the sensitivity of 
complete sets of polarization transfer observables to reductions of the meson 
masses and meson-nucleon coupling constants varying from 0\% to 20\%: the 
vertically hatched and dotted bands in Figs.~(\ref{fig:fig5}), (\ref{fig:fig6}) 
and (\ref{fig:fig7}) represent the sensitivity of a particular spin observable 
to reductions of coupling constants and meson masses ranging from 0\% to 20\% 
for both FR-DWIA and ZR-DWIA models, respectively.

For the knockout from all three states, we see that A$_{y}$, P and D$_{s' s}$
are very sensitive to reductions in the meson-coupling constants and meson masses. 
On the other hand, the spin observables D$_{s' \ell}$ and D$_{\ell ' \ell}$ 
exhibit minimal sensitivity to nuclear medium effects. Note that at the point
corresponding to minimum recoil momentum, the spin observables D$_{n n}$,
D$_{s' \ell}$ and D$_{\ell' \ell}$ are insensitive to nuclear medium effects
for 3s$_{1/2}$ knockout. On the other hand, for both 2d$_{3/2}$ and 2d$_{5/2}$ 
states, D$_{s' \ell}$  and D$_{\ell' \ell}$ also exhibit minimal sensitivity 
to nuclear medium corrections at minimum recoil. This is also the case for 
D$_{n n}$ and D$_{\ell' \ell}$ for the 2d$_{5/2}$ state.

\section{\label{sec:summary-and-conclusions}Summary and conclusions}
In this paper we have exploited the discriminatory nature of 
complete sets of polarization transfer observables (P, A$_{y}$, D$_{n n}$, 
D$_{s' s}$, D$_{s' \ell}$, D$_{\ell' s}$ and D$_{\ell' \ell}$) for exclusive 
($\vec{p},2 \vec{p}\,$) reactions to address a number of important physics issues.
One of our aims was to identify specific observables which can yield information
on whether the relativistic Dirac equation or the nonrelativistic Schr\"{o}dinger
equation is the more appropriate dynamical equation for the description of polarization
phenomena within the framework of distorted wave impulse-approximation models.
In addition, we also studied the sensitivity of spin observables to nuclear 
distortion effects, finite-range (FR) versus zero-range (ZR) approximations of 
the relativistic DWIA, as well as to reductions of meson-nucleon coupling constants 
and meson masses by the surrounding nuclear medium in which the NN interaction occurs. 
In particular, we focused on proton knockout from the 3s$_{1/2}$, 2d$_{3/2}$ 
and 2d$_{5/2}$ states in $^{208}$Pb, at an incident laboratory kinetic energy 
of 202 MeV, and for coincident coplanar scattering angles ($28.0^{\circ}$, $-54.6^{\circ}$).
The motivation for choosing a heavy target nucleus, $^{208}$Pb, and a relatively low 
incident energy of 202 MeV is to maximize the influence of distortion effects as well 
as maximize differences between FR and ZR approximations to the relativistic DWIA,
while still maintaining the validity of the impulse approximation, and also avoiding 
complications associated with the inclusion of recoil corrections in the relativistic 
Dirac equation. Another important consideration for our choice of reaction kinematics
is the availability of analyzing power data to provide initial constraints on current 
distorted wave models. Unfortunately, there are no published data on other spin observables 
for the reaction kinematics of interest. 

Previously, we established the clear superiority of relativistic DWIA models, 
compared to the nonrelativistic DWIA models, for describing exclusive ($\vec{p},2 p$) 
analyzing powers \cite{Hi03}. In this paper, we identify additional observables which display large 
differences to Dirac- versus Schr\"{o}dinger-equation-based models and which need to be 
measured in order to check the consistency of the analyzing power predictions regarding 
the role of different dynamical models. In particular, for knockout from the 3s$_{1/2}$ 
state the spin observables A$_{y}$, D$_{n n}$ and D$_{s' s}$ exhibit large differences 
between Dirac and Schr\"{o}dinger-based DWIA models. On the other hand, for knockout 
from the 2d$_{3/2}$ state the most sensitive observables to dynamical differences are 
D$_{n n}$, D$_{\ell ' s}$ and D$_{\ell ' \ell}$, whereas for 2d$_{5/2}$ knockout the  
corresponding observables are D$_{s' s}$ and D$_{\ell ' s}$.

Regarding observables which display large differences between FR and ZR
approximations to the relativistic DWIA, we see that for knockout from the 
3s$_{1/2}$ state, the induced polarization, P, is the most sensitive, whereas
D$_{\ell' s}$  and D$_{s' \ell}$ are the most sensitive observables for 2d$_{3/2}$ 
and 2d$_{5/2}$ knockout, respectively. 

We have also established that all polarization transfer observables are 
relatively insensitive to different global Dirac optical potential parameter 
sets. In addition, by comparing relativistic DWIA predictions to corresponding
plane wave predictions, we also demonstrated the importance of distorting potentials 
for describing the oscillatory behavior of spin observables thus refuting, for the 
first time, qualitative arguments that spin observables are insensitive to nuclear 
distortion effects.

We have also shown that the analyzing power data alone are unable to establish whether 
the nuclear medium does indeed reduce meson-nucleon coupling constants and meson masses: 
on one hand, the relativistic ZR predictions suggest that the scattering matrix for NN 
scattering in the nuclear medium is adequately represented by the corresponding matrix for 
free NN scattering. On the other hand, the relativistic FR results suggest that a 
10\% to 20\% reduction of meson-nucleon coupling constants and meson masses by the nuclear 
medium is essential for providing a consistent description of the 3s$_{1/2}$, 2d$_{3/2}$ 
and 2d$_{5/2}$ analyzing powers \cite{Hi03}. In this paper we studied the sensitivity
of the other polarization transfer observables to reductions in these parameters varying 
between 0\% and 20\%. For the knockout from all three states we see that A$_{y}$, P and 
D$_{s' s}$ are very sensitive to reductions in the meson-coupling constants and meson masses. 

We also established a number of interesting model predictions for spin observables 
at the kinematic point corresponding to minimum recoil momentum. For 3s$_{1/2}$ 
knockout the relativistic plane wave, ZR-DWIA and nonrelativistic DWIA predictions 
are virtually identical for the induced polarization (P), D$_{s' \ell}$ and D$_{\ell ' s}$; 
ZR-DWIA and FR-DWIA predictions are nearly identical for both D$_{n n}$ and 
D$_{s ' \ell}$. For knockout from the 2d$_{3/2}$ state both FR-DWIA and ZR-DWIA 
yield similar results for P, D$_{s' s}$ and D$_{s' \ell}$; P and D$_{s' s}$ 
are insensitive to relativistic nuclear distortion at the point in question. We also observe 
that for 2d$_{5/2}$ knockout, relativistic plane wave, FR-DWIA and nonrelativistic 
DWIA predictions are virtually identical for D$_{s' \ell}$ and D$_{\ell' s}$. 
Regarding the influence of nuclear medium effects, the spin observables D$_{n n}$, 
D$_{s' \ell}$ and D$_{\ell' \ell}$ are insensitive for 3s$_{1/2}$ knockout. On the 
other hand, for the 2d$_{3/2}$ and 2d$_{5/2}$ states, D$_{s' \ell}$  and D$_{\ell' \ell}$ 
also exhibit minimal sensitivity to nuclear medium corrections at minimum recoil. This 
is also the case for D$_{n n}$ and D$_{\ell' \ell}$ for the 2d$_{5/2}$ state. Hence, 
by measuring spin observables at minimum recoil momentum one can eliminate differences 
between different dynamical models and model parameters and focus on a specific issue 
of interest.

Once again, we stress the urgent need for experimental data on polarization observables,
in addition to the commonly-measured analyzing power, in order to resolve issues concerning
the role of relativistic versus nonrelativistic dynamics in nuclear physics, as well as study
the influence of the  nuclear medium on the strong interaction.

\begin{acknowledgements}
G.C.H acknowledges financial support from the Japanese 
Ministry of Education, Science and Technology for research conducted 
at the Research Center for Nuclear Physics, Osaka University, Osaka, 
Japan. This material is based upon work supported by the National
Research Foundation under Grant numbers: GUN 2058507 (J.M), 2053786 (G.C.H).
\end{acknowledgements}

\end{document}